\input harvmac

\noblackbox
\input epsf


\mathchardef\varGamma="0100
\mathchardef\varDelta="0101
\mathchardef\varTheta="0102
\mathchardef\varLambda="0103
\mathchardef\varXi="0104
\mathchardef\varPi="0105
\mathchardef\varSigma="0106
\mathchardef\varUpsilon="0107
\mathchardef\varPhi="0108
\mathchardef\varPsi="0109
\mathchardef\varOmega="010A

\font\mbm = msbm10
\font\Scr=rsfs10 
\def\bb#1{\hbox{\mbm #1}}

\def\scr#1{\hbox{\Scr #1}}

\def\Mt{{\kern1em\hbox{$\tilde{\kern-1em{\scr M}}$}}}
\def\At{{\kern1em\hbox{$\tilde{\kern-1em{\scr A}}$}}}
\def\Kt{{\kern1em\hbox{$\tilde{\kern-1em{\scr K}}$}}}

\font\sScr=rsfs7 
\def\sscr#1{\hbox{\sScr #1}}


\nref\GiddingsYU{
S.~B.~Giddings, S.~Kachru and J.~Polchinski,
{\it Hierarchies from fluxes in string compactifications,}
Phys.\ Rev.\ D66 (2002) 106006
[arXiv:hep-th/0105097].
}

\nref\KachruAW{
S.~Kachru, R.~Kallosh, A.~Linde and S.~P.~Trivedi,
{\it De Sitter vacua in string theory,}
Phys.\ Rev.\ D68 (2003) 046005
[arXiv:hep-th/0301240].
}

\nref\vonGersdorffBF{
G.~von Gersdorff and A.~Hebecker,
{\it Kaehler corrections for the volume modulus of flux compactifications,}
Phys.\ Lett.\ B624 (2005) 270
[arXiv:hep-th/0507131].
}

\nref\BergJA{
M.~Berg, M.~Haack and B.~Kors,
{\it String loop corrections to Kaehler potentials in orientifolds,}
JHEP 0511 (2005) 030
[arXiv:hep-th/0508043].
}

\nref\BergYU{
M.~Berg, M.~Haack and B.~Kors,
{\it On volume stabilization by quantum corrections,}
Phys.\ Rev.\ Lett.\  96 (2006) 021601
[arXiv:hep-th/0508171].
}

\nref\IrgesAU{
N.~Irges,
{\it Anomalous U(1), holomorphy, supersymmetry breaking and dilaton stabilization,}
Phys.\ Rev.\ D59 (1999) 115008
[arXiv:hep-ph/9812338].
}

\nref\AntoniadisPP{
I.~Antoniadis and T.~Maillard,
{\it Moduli stabilization from magnetic fluxes in type I string theory,}
Nucl.\ Phys.\ B716 (2005) 3
[arXiv:hep-th/0412008].
}

\nref\AntoniadisNU{
I.~Antoniadis, A.~Kumar and T.~Maillard,
{\it Moduli stabilization with open and closed string fluxes,}
arXiv:hep-th/0505260.
}

\nref\BianchiYZ{
M.~Bianchi and E.~Trevigne,
{\it The open story of the magnetic fluxes,}
JHEP 0508 (2005) 034
[arXiv:hep-th/0502147].
}

\nref\BianchiSA{
M.~Bianchi and E.~Trevigne,
{\it Gauge thresholds in the presence of oblique magnetic fluxes,}
JHEP 0601 (2006) 092
[arXiv:hep-th/0506080].
}

\nref\KumarER{
A.~Kumar, S.~Mukhopadhyay and K.~Ray, 
{\it Moduli stabilization with non-Abelian fluxes,}
arXiv:hep-th/0605083.
}

\nref\GinspargWR{
P.~H.~Ginsparg and C.~Vafa,
{\it Toroidal Compactification Of Non-Supersymmetric Heterotic Strings,}
Nucl.\ Phys.\ B289 (1987) 414.
}

\nref\NairZN{
V.~P.~Nair, A.~D.~Shapere, A.~Strominger and F.~Wilczek,
{\it Compactification Of The Twisted Heterotic String,}
Nucl.\ Phys.\ B287 (1987) 402.
}

\nref\ItzhakiZR{
N.~Itzhaki, D.~Kutasov and N.~Seiberg,
{\it Non-supersymmetric deformations of non-critical superstrings,}
JHEP 0512 (2005) 035
[arXiv:hep-th/0510087].
}

\nref\HarmarkSF{
T.~Harmark, V.~Niarchos and N.~A.~Obers,
{\it Stable non-supersymmetric vacua in the moduli space of non-critical
superstrings,}
arXiv:hep-th/0605192.
}

\nref\ScherkTA{
J.~Scherk and J.~H.~Schwarz,
{\it Spontaneous Breaking Of Supersymmetry Through Dimensional Reduction,}
Phys.\ Lett.\ B82 (1979) 60.
}

\nref\ScherkZR{
J.~Scherk and J.~H.~Schwarz,
{\it How To Get Masses From Extra Dimensions,}
Nucl.\ Phys.\ B153 (1979) 61.
}

\nref\SSstringi{
R.~Rohm,
{\it Spontaneous Supersymmetry Breaking In Supersymmetric String
Theories,}
Nucl.\ Phys.\ B237 (1984) 553.
}

\nref\SSstringii{
C.~Kounnas and M.~Porrati,
{\it Spontaneous Supersymmetry Breaking In String Theory,}
Nucl.\ Phys.\ B310 (1988) 355.
}

\nref\FKPZ{
S.~Ferrara, C.~Kounnas, M.~Porrati and F.~Zwirner,
{\it Superstrings With Spontaneously Broken Supersymmetry And Their 
Effective Theories,}
Nucl.\ Phys.\ B318 (1989) 75;
}

\nref\SSstringiii{
C.~Kounnas and B.~Rostand,
{\it Coordinate dependent compactifications and discrete symmetries,}
Nucl.\ Phys.\ B341 (1990) 641.
}

\nref\elias{
E.~Kiritsis and C.~Kounnas,
{\it Perturbative and non-perturbative partial supersymmetry breaking:  $N = 4
\to N = 2 \to N = 1$,}
Nucl.\ Phys.\ B503 (1997) 117
[arXiv:hep-th/9703059].
}

\nref\McClainID{
B.~McClain and B.~D.~B.~Roth,
{\it Modular Invariance For Interacting Bosonic Strings At Finite Temperature,}
Commun.\ Math.\ Phys.\  111 (1987) 539.
}

\nref\OBrienPN{
K.~H.~O'Brien and C.~I.~Tan,
{\it Modular Invariance Of Thermopartition Function And Global Phase Structure
Of Heterotic String,}
Phys.\ Rev.\ D36 (1987) 1184.
}

\nref\ItoyamaEI{
H.~Itoyama and T.~R.~Taylor,
{\it Supersymmetry Restoration In The Compactified $O(16)\times O'(16)$ Heterotic String Theory,}
Phys.\ Lett.\ B186 (1987) 129.
}

\nref\DixonPC{
L.~J.~Dixon, V.~Kaplunovsky and J.~Louis,
{\it Moduli dependence of string loop corrections to gauge coupling constants,}
Nucl.\ Phys.\ B355 (1991) 649.
}

\nref\MayrMQ{
P.~Mayr and S.~Stieberger,
{\it Threshold corrections to gauge couplings in orbifold compactifications,}
Nucl.\ Phys.\ B407 (1993) 725  [arXiv:hep-th/9303017].
}

\nref\GhilenceaBV{
D.~M.~Ghilencea, H.~P.~Nilles and S.~Stieberger,
{\it Divergences in Kaluza-Klein models and their string regularization,}
New J.\ Phys.\  4 (2002) 15
[arXiv:hep-th/0108183].
}

\nref\TraplettiUK{
M.~Trapletti,
{\it On the unfolding of the fundamental region in integrals of modular
invariant amplitudes,}
JHEP 0302 (2003) 012
[arXiv:hep-th/0211281].
}

\nref\HagedornST{
R.~Hagedorn,
{\it Statistical thermodynamics of strong interactions at high-energies,}
Nuovo Cim.\ Suppl.\  3 (1965) 147.
}

\nref\KiritsisEN{
E.~Kiritsis, C.~Kounnas, P.~M.~Petropoulos and J.~Rizos,
{\it String threshold corrections in models with spontaneously broken supersymmetry,}
Nucl.\ Phys.\ B540 (1999) 87
[arXiv:hep-th/9807067].
}

\nref\AngelantonjID{
C.~Angelantonj, E.~Dudas and J.~Mourad,
{\it Orientifolds of string theory Melvin backgrounds,}
Nucl.\ Phys.\ B637 (2002) 59
[arXiv:hep-th/0205096].
}

\nref\DudasVI{
E.~Dudas and C.~Timirgaziu,
{\it Non-tachyonic Scherk-Schwarz compactifications, cosmology and moduli stabilization,}
JHEP 0403 (2004) 060
[arXiv:hep-th/0401201].
}

\nref\KiritsisHF{
E.~Kiritsis and N.~A.~Obers,
{\it Heterotic/type-I duality in $D < 10$ dimensions, threshold corrections  and
D-instantons,}
JHEP 9710 (1997) 004
[arXiv:hep-th/9709058].
}

\nref\SugimotoTX{
S.~Sugimoto,
{\it Anomaly cancellations in type I $D9-\bar{D9}$ system and the USp(32)  
string theory,}
Prog.\ Theor.\ Phys.\ 102 (1999) 685
[arXiv:hep-th/9905159];
}

\nref\Ads{
I.~Antoniadis, E.~Dudas and A.~Sagnotti,
{\it Brane supersymmetry breaking,}
Phys.\ Lett.\ B464 (1999) 38
[arXiv:hep-th/9908023];
}

\nref\AngelantonjJH{
C.~Angelantonj,
{\it Comments on open-string orbifolds with a non-vanishing $B_{ab}$,}
Nucl.\ Phys.\ B566 (2000) 126
[arXiv:hep-th/9908064].
}

\nref\Alda{
G.~Aldazabal and A.~M.~Uranga,
{\it Tachyon-free non-supersymmetric type IIB orientifolds via  
brane-antibrane systems,}
JHEP 9910 (1999) 024
[arXiv:hep-th/9908072].
}

\nref\AtickSI{
J.~J.~Atick and E.~Witten,
{\it The Hagedorn Transition and the Number of Degrees of Freedom of String Theory,}
Nucl.\ Phys.\ B310 (1988) 291.
}

\nref\AntoniadisKH{
I.~Antoniadis and C.~Kounnas,
{\it Superstring phase transition at high temperature,}
Phys.\ Lett.\ B261 (1991) 369.
}

\nref\PolchinskiZF{
J.~Polchinski,
{\it Evaluation Of The One Loop String Path Integral,}
Commun.\ Math.\ Phys.\  104 (1986) 37.
}

\nref\DitsasPM{
P.~Ditsas and E.~G.~Floratos,
{\it Finite Temperature Closed Bosonic String in a Finite Volume,}
Phys.\ Lett.\ B201 (1988) 49.
}

\nref\OsorioHI{
M.~A.~R.~Osorio and M.~A.~Vazquez-Mozo,
{\it Strings Below The Planck Scale,}
Phys.\ Lett.\ B280 (1992) 21
[arXiv:hep-th/9201044].
}

\nref\OsorioVG{
M.~A.~R.~Osorio and M.~A.~Vazquez-Mozo,
{\it Duality in nontrivially compactified heterotic strings,}
Phys.\ Rev.\ D47 (1993) 3411
[arXiv:hep-th/9207002].
}

\nref\DavisQE{
J.~L.~Davis, F.~Larsen and N.~Seiberg,
{\it Heterotic strings in two dimensions and new stringy phase transitions,}
JHEP 0508 (2005) 035
[arXiv:hep-th/0505081].
}

\nref\DienesDU{
K.~R.~Dienes and M.~Lennek,
{\it Adventures in thermal duality. I: Extracting closed-form solutions for
finite-temperature effective potentials in string theory,}
Phys.\ Rev.\ D70 (2004) 126005
[arXiv:hep-th/0312216].
}

\nref\DienesDV{
K.~R.~Dienes and M.~Lennek,
{\it Adventures in thermal duality. II: Towards a duality-covariant string
thermodynamics,}
Phys.\ Rev.\ D70 (2004) 126006
[arXiv:hep-th/0312217].
}

\nref\DienesVW{
K.~R.~Dienes and M.~Lennek,
{\it Re-identifying the Hagedorn transition,}
arXiv:hep-th/0505233.
}

\nref\DienesXA{
K.~R.~Dienes and M.~Lennek,
{\it How to extrapolate a string model to finite temperature: Interpolations and
implications for the Hagedorn transition,}
arXiv:hep-th/0507201.
}


\Title{\vbox{
\rightline{\tt hep-th/0608022} 
\rightline{DFTT 16/2006}
\rightline{IFUM 869-FT}
}}
{\vbox{
\centerline{An Alternative for Moduli Stabilisation}
}}

\centerline{Carlo Angelantonj$^\dagger$,
Matteo Cardella$^\ddagger$  and Nikos Irges$^\star$
}
\medskip
\centerline{\it $^\dagger$ Dipartimento di Fisica Teorica, Universit\`a di Torino,}
\centerline{\it INFN Sezione di Torino, Via P. Giuria 1, 10125 Torino, Italy}
\centerline{\it $^\ddagger$ Dipartimento di Fisica dell'Universit\`a di Milano,}
\centerline{\it INFN sezione di Milano, via Celoria 16, 20133 Milano, Italy}
\centerline{\it $^\star$ High Energy and Elementary Particle Physics Division,}
\centerline{\it Department of Physics, University of Crete, 71003 Heraklion, Greece}

\vskip 0.6in

\centerline{\bf Abstract}

\noindent
The one-loop vacuum energy is explicitly computed for a class of perturbative string vacua where supersymmetry is spontaneously broken by a T-duality invariant asymmetric Scherk-Schwarz deformation. The low-lying spectrum is tachyon-free for any value of the compacti\-fication radii and thus no Hagedorn-like phase-transition takes place. Indeed, the induced effective potential is free of divergence, and has a global anti de Sitter minimum where geometric moduli are naturally stabilised.

\Date{July, 2006}

The true vacuum state in a quantum field theory is given by the configuration with the lowest vacuum energy. In general, superstrings yield a vanishing vacuum energy for toroidal or orbifold compactifications if supersymmetry is intact.  As a consequence, superstring compactifications are usually characterised by a moduli space of supersymmetric vacua. This moduli space is spanned by certain dynamical moduli with vanishing potential, whose undetermined vacuum expectation values fix the shape and size of compact internal spaces, and set the strength of gravitational and gauge interactions. This is clearly an embarrassment for any attempt at phenomenological study in String Theory since experiments set severe bounds on Brans-Dicke-like forces and on time variation of coupling constants. It is then clear that the search for mechanisms for moduli stabilisation in string compactifications is of utmost importance. Recently, considerable progress has been made along this direction by allowing for non-trivial fluxes. In this class of compactifications, the internal manifold is permeated by constant fluxes for the field strengths of some Neveu-Schwarz--Neveu-Schwarz and Ramond--Ramond fields. In this way, a non-trivial potential is generated whose extrema fix the vacuum expectation values of complex structure mo\-du\-li as well as the dilaton field \GiddingsYU. Moreover, the inclusion of non-perturbative effects (like gaugino condensation) and/or the emergence of perturbative string-loop and $\alpha'$ corrections leads also to the stabilisation of K\"ahler class moduli 
\refs{\KachruAW{--}\BergYU}\footnote{$^\ast$}{See \IrgesAU\ for an earlier attempt at stabilising the dilaton using perturbative $g_{\rm s}$ and $\alpha '$ corrections.}. 
Despite the indiscussed importance of these results, any attempt to lift the moduli space of supersymmetric vacua via geometric fluxes and non-perturbative effects relies on a low-energy supergravity analysis, while a full-fledged perturbative string description is by definition missing. Alternative approaches to the moduli stabilisation problem involve the introduction of open-string magnetic backgrounds \refs{\AntoniadisPP{--}\KumarER}. Although a perturbative string description is now in principle available (at least for Abelian backgrounds), no consistent string-theory vacua with stabilised moduli have been constructed so far. 

An alternative way of stabilising moduli might instead rely on non-supersymmetric string compactifications where corrections to the moduli space of supersymmetric vacua are generated in perturbation theory. This approach has the great advantage of allowing 
a perturbative string theory description, and can be democratically realised in heterotic strings, type II superstrings and their orientifolds, both on toroidal and orbifold backgrounds. 
In this case, the first correction to the flat potential for moduli fields is determined by the non-vanishing one-loop vacuum energy. Despite non-supersymmetric strings are typically plagued by the presence of tachyons in the twisted sector, thus inducing a divergent vacuum energy, tachyon-free heterotic models have been constructed in the past and it has also been shown that the associated one-loop vacuum energies are finite and have extrema at symmetry-enhancement points \refs{\GinspargWR, \NairZN}. In this letter, we extend this analysis to superstrings with spontaneously broken supersymmetry and show how, in models without tachyons, symmetries of the (deformed) Narain lattice determine the minima of the one-loop cosmological constant. Constructions with a similar target have been carried out recently also in the context of non-critical string theory 
\refs{\ItzhakiZR,\HarmarkSF}.

Among the various mechanisms for breaking supersymmetry, the Scherk-Schwarz deformation provides an elegant realisation based on compactification\refs{\ScherkTA, \ScherkZR}. In the simplest case of circle compactification, it amounts in Field Theory to allowing the higher dimensional fields to be periodic around the circle up to an R-symmetry transformation. The Kaluza-Klein momenta of the various fields are correspondingly shifted proportionally to their R charges, and modular invariance dictates the extension of this mechanism to the full perturbative spectrum in models of oriented closed strings 
\refs{\SSstringi{--}\SSstringiii}. Actually, it is a known fact that Scherk-Schwarz deformations can be conveniently realised in String Theory as freely acting orbifolds \elias. Typically, non-supersymmetric projections are accompanied by shifts of internal coordinates, so that the moduli describing their size set the supersymmetry-breaking scale. The simplest instance of a Scherk-Schwarz deformation involves the $\bb{Z}_2$ orbifold generated by $g=(-1)^F \, \delta$, where $(-1)^F$ is the space-time fermion index and $\delta$ shifts the compact coordinate $y\in S^1 (R)$ by half of the length of the circle, $\delta: \ y \to y + \pi R$. As usual, the resulting string spectrum is encoded in the one-loop partition function 
$$
\eqalign{
{\scr Z} = -  {V_9 \over 2 (4\pi^2 \alpha ')^{9/2}}& \int_{\sscr F} {d^2\tau \over \tau_2^{11/2}} \, \sum_{m,n} \left\{\left[ \left| {V_8 - S_8 \over \eta^8}\right|^2 + \left| {V_8 + S_8 \over \eta^8}\right|^2 \, (-1)^m \right]\, \varLambda_{m,n} (R)\right.
\cr
&+ \left. \left[ \left| {O_8 - C_8 \over \eta^8}\right|^2 + \left| {O_8 + C_8 \over \eta^8}\right|^2 \, (-1)^m \right]\, \varLambda_{m,n+{1\over 2}} (R)\right\} \,,
\cr}
\eqno(1)
$$
where $V_9$ is the (infinite) volume of the non-compact dimensions, 
$(O_8, V_8, S_8 , C_8)$ are the characters associated with the SO(8) little group, and 
$$
\varLambda_{m+a,n+b} (R) = q^{{\alpha'\over 4} \left( {m+a\over R} + {(n+b)R\over \alpha '}\right)^2} \, \bar q^{{\alpha'\over 4} \left( {m+a\over R} - {(n+b)R\over \alpha '}\right)^2}
=: q^{{\alpha'\over 4} p^2_{\rm L}} \, \bar q^{{\alpha'\over 4} p^2_{\rm R}}
$$ 
encodes the contribution of momentum and winding zero-modes.  
While the first line clearly spells-out the mass-shift between bosons and fermions induced by the Scherk-Schwarz deformation, the presence of states with reversed GSO projection (the second line) is crucial for modular invariance of ${\scr Z}$. These states have masses
$$
m_{\rm tw}^2 = -{1\over \alpha '} + {1\over 2} \left({m\over R}\right)^2 + {1\over 2} \left( {(n+{\textstyle{1\over 2}} )R \over \alpha '}\right)^2 + {\rm oscillators} \eqno(2)
$$
and thus the tachyonic ground state is actually massive for large values of $R$, while it is massless and then really tachyonic for\footnote{$^\dagger$}{Actually, to compare with the standard Scherk-Schwarz deformation one has to halve the compactification radius, so that bosons and fermions have indeed integer and half-odd-integer KK momenta, and the twisted tachyon appears for $R<\sqrt{2\alpha '}$.} $R\le \sqrt{8\alpha '}$.

The absence of tachyonic excitations in the large-radius regime suggests that in this range the theory is perturbatively under control. Using standard techniques
\refs{\McClainID{--}
\TraplettiUK}, one can unfold the fundamental domain ${\scr F}= \{ |\tau | \ge 1 \,, \ -{1\over 2} < \tau_1 \le {1\over 2} \}$ into the half-infinite strip ${\scr S} =\{ -{1\over 2} < \tau_1 \le {1\over 2}\,,\ 0\le \tau_2 < +\infty \}$ to compute the one-loop  potential
$$
{\scr V}\, (R)= - {7936 \, \pi^5\over 945}\, \left({\sqrt{\alpha'}\over R}\right)^9  
 - 2^5\, \left({\sqrt{\alpha '}\over R}\right)^4 \sum_{p\, {\rm odd}} \sum_{N=1}^\infty \, d_N^2 
{N^{5/2}\over p^5}\, K_5 \left( 2\pi p \sqrt{N R^2/\alpha'}\right)\,, 
$$
where $d_N$ counts the degeneracy of states at each mass level, $V_8/\eta^8 = \sum_{N=0}^\infty d_N q^N$. The contribution of massless-states reproduces the standard field theory result, while massive-state contributions are exponentially suppressed for large values of $R$, but yield a divergent contribution in the tachyonic region $R\le \sqrt{8\alpha'}$.
This divergence is a consequence of the exponential growth of string states 
$$
d_N \sim N^{-11/4} e^{+\pi \sqrt{8N}} \,, 
$$
and is responsible for the well-known first-order Hagedorn phase transition \HagedornST, after we reinterpret the compact radius in terms of the finite temperature, $\beta \sim R$. 

Although this representation of the Scherk-Schwarz deformation has a natural field theory limit (after the compactification radius is properly halved), where the anti-periodic fermions have half-integer Kaluza-Klein excitations and the vacuum energy has the typical $R^{-n}$ behaviour, for $n$ non-compact dimensions, String Theory can afford more possibilities:
one can actually deform only the momenta, as in eq. (1), only the windings or both. These correspond in turn to the freely acting orbifolds $g\, \delta_i$, where $g$ is any non-supersymmetric generator and the $\delta_i$ act as
$$
\delta_1 = \left\{ 
\eqalign{
X_{\rm L} &\to X_{\rm L} + {\pi R\over 2}
\cr
X_{\rm R} &\to X_{\rm R} + {\pi R\over 2}
\cr}\right.\,,
\qquad
\delta_2 = \left\{ 
\eqalign{
X_{\rm L} &\to X_{\rm L} + {\alpha '\pi \over 2R}
\cr
X_{\rm R} &\to X_{\rm R} - {\alpha ' \pi \over 2R}
\cr}\right.\,,
$$
and
$$
\delta_3 = \left\{
\eqalign{
X_{\rm L} &\to X_{\rm L} + {\pi R\over 2}+ {\alpha '\pi \over 2R}
\cr
X_{\rm R} &\to X_{\rm R} + {\pi R\over 2}- {\alpha ' \pi \over 2R}
\cr}\right.
$$
on a circle of radius $R$.

Clearly the $\delta_2$ shift does not introduce new physics, since it is directly related to the  $\delta_1$ shift via T-duality: $R\to \alpha ' /R$. As a result, it yields a tachyon-free spectrum in the small-radius regime, the associated effective potential behaves like $R^n$, while a Hagedorn-like transition occurs now at $R=\sqrt{\alpha'/8}$. 

More interesting is instead the $\delta_3$ shift. It preserves T-duality and thus one can expect to have a sensible theory both in the small-radius and in the large-radius 
regimes\footnote{$^\star$}{Actually, this asymmetric $\delta_3 \, (-1)^F$ freely acting orbifold with diagonal metric is equivalent to a more conventional symmetric Scherk-Schwarz deformation but with a non-vanishing $B$-field background, as shown in \KiritsisEN. However, in the following we concentrate on the asymmetric shift since it is clearly T-duality invariant and thus some results are easier to prove.}. In this case, a generic vertex operator $V_{m,n} = e^{ip_{\rm L} X_{\rm L} + i p_{\rm R} X_{\rm R}}$, with $p_{\rm L,R} = {m\over R} \pm {nR\over \alpha '}$, gets the phase $(-1)^{m+n}$ under the action of $\delta_3$, and, as a consequence, both momenta and windings are half-odd-integers in the twisted sector. In this sector the mass formulae read
$$
\eqalign{
m^2_{\rm L} =& -{1\over 2\alpha '} + {\alpha ' \over 4} \sum_i \left( 
{m_i+{1\over 2}\over R_i} +{(n_i+{\textstyle{1\over 2}} )R_i \over \alpha '}\right)^2 + N^{(X)} + N^{(\psi)} \,,
\cr
m^2_{\rm R} =& -{1\over 2\alpha '} + {\alpha ' \over 4} \sum_i \left( 
{m_i+{1\over 2}\over R_i} -{(n_i+{\textstyle{1\over 2}} )R_i \over \alpha '}\right)^2 + \tilde N^{(X)} + \tilde N^{(\psi)} \,,
\cr}
$$ 
where we have consider the more general case of a higher-dimensional internal manifold consisting of the product of various circles, each of radius $R_i$. 
Evidently, level-matching 
$$
N^{(X)} + N^{(\psi)} - \tilde N^{(X)} - \tilde N^{(\psi)} + \left( m+{\textstyle{1\over 2}} \right) \cdot
\left( n+ {\textstyle{1\over 2}}\right) =0
$$
is not satisfied if $\delta_3$ acts on a single coordinate. Actually this deformation is only allowed when acting on coordinates of an even $2d$-dimensional torus. In this case, the lightest states have masses
$$
\alpha '\,m^2_{\rm tw} = {\textstyle{1\over 8}} \sum_{i=1}^{2d} \left( {\sqrt{\alpha '}\over R} - {R\over \sqrt{\alpha'}}\right)^2 +{d-2\over 2}\,,
$$
and the twisted spectrum is clearly free of tachyons for $d=2,3$\footnote{$\ddagger$}{It is amusing to note similarities with mass-formulae of other tachyon-free non-supersymmetric orientifold models \refs{\AngelantonjID,\DudasVI}}. Henceforth, one expects to have a finite and well-behaved one-loop result for any values of $R$. This is in contrast to the previous case, where a divergence induced by the emergence of tachyonic excitations triggers a first-order phase transition. 
In fact, in the case at hand the partition function schematically reads
$$
{\scr Z} = - {V_{10-2d}\over 2 (4\pi^2 \alpha')^{5-d}} \int_{\sscr F} {d^2\tau \over \tau_2^{6-d}}\, \sum_{\{m^2\}} \, c (m^2) \, q^{m^2_{\rm L}} \bar q ^{m^2_{\rm R}}
$$
and thus the absence of tachyons prevents from IR divergence while, as usual, modular invariance excludes the dangerous UV region from the integration domain ${\scr F}$.

To be more precise, the complete partition function now reads
$$
\eqalign{
{\scr Z} = - {V_{10-2d}\over 2 (4\pi^2 \alpha')^{5-d}} &\int_{\sscr F} {d^2\tau \over \tau_2^{6-d}}\, \sum_{\{m,n\}} \left\{ \left[ \left| { V_8 - S_8 \over \eta^8}\right|^2 + 
\left| { V_8 + S_8 \over \eta^8}\right|^2 \, (-1)^{(m+n)\cdot\epsilon} \right] \, \varLambda_{m,n} (R_i)\right.
\cr
&+\left. \left[ \left|{ O_8 - C_8 \over \eta^8}\right|^2 + 
\left| { O_8 + C_8 \over \eta^8}\right|^2 \, (-1)^{d+(m+n)\cdot\epsilon} \right] \, \varLambda_{m+{1\over 2},n+{1\over 2}} (R_i)\right\}
\cr}
$$
where $\epsilon = (1,1,\ldots, 1)$ is a $2d$-dimensional unit vector, and the second line clearly spells-out the ``non-canonical'' deformation of the Narain lattice in the twisted sector.

Also in this case, one can use standard unfolding techniques to convert the integral over the fundamental domain into the integral over the half-infinite strip. Typically, this procedure involves a Poisson re-summation in order to disentangle the contributions to the integral of different orbits of ${\rm SL} (2,\bb{Z})$. Re-summing over windings or momenta clearly spells-out the small or large radius behaviour. For this reason, in the case of $\delta_1$ and $\delta_2$ shifts the very consistency of the perturbative string expansion suggests to re-sum over momenta and windings, respectively, while in the more interesting case of a $\delta_3$ shift either ways are meaningful, since the spectrum is free of tachyons and no first-order phase transitions are expected to occur. In particular, from
$$
{\scr V} = - {1\over 2(4\pi^2 \alpha')^{5-d} } \int_{\tilde{\!\sscr F}} {d^2\tau\over \tau_2^{6-d}} \, \left| {\theta_2^4 (0|\tau) \over \eta^{12} (\tau)}\right|^2 \sum_{\{m,n\} } 
 (-1)^{(m+n)\cdot\epsilon}\, \varLambda_{m,n}
$$
with ${\kern.5em\hbox{$\tilde{\kern-.5em{\scr F}}$}} = (1 + S + ST ) \circ {\scr F}$, Poisson re-summations over the $2d$ momenta or windings yield
$$
{\scr V}_{\rm large} = - {\prod_i R_i \over 2(4\pi^2 )^{5-d} (\alpha ')^5} \int_{\tilde{\!\sscr F}} {d^2\tau\over \tau_2^{6}} \, \left| {\theta_2^4 \over \eta^{12}}\right|^2 
\sum_{\{\ell,n\} } e^{i\pi n\cdot \epsilon}\, \exp\left\{ -{\pi \over 4\alpha ' \tau_2} \left\| \left(2\ell +1 + 2n \tau\right) R \right\|^2 \right\} \,,
$$
and
$$
{\scr V}_{\rm small} = - {\prod_i R_i^{-1} \over 2(4\pi^2 )^{5-d} (\alpha ')^{5-2d}} \int_{\tilde{\!\sscr F}} {d^2\tau\over \tau_2^{6}} \, \left| {\theta_2^4 \over \eta^{12}}\right|^2 
\sum_{\{m,k\} } e^{i\pi m\cdot \epsilon}\, \exp\left\{ -{\pi \alpha ' \over 4 \tau_2} \left\| {2k +1 + 2m \tau\over R} \right\|^2 \right\} \,,
$$
respectively, where $\|a\|^2 = \sum_{i=1}^{2d} a^\dagger_i a_i$. An element $A$ of $\varGamma_0[2]$ acts as left multiplication on the  $(2\times 2d)$-dimensional integral matrices
$$
M= \left( \matrix{ 2n_1 &  \ldots & 2n_{2d} \cr
2\ell_1 +1 &  \ldots & 2\ell_{2d}+1 \cr}\right)\,,\qquad
\tilde M = \left( \matrix{ 2m_1 & \ldots & 2m_{2d} \cr
2k_1 +1 & \ldots & 2k_{2d} +1\cr}\right) \,.
$$
As a result, both $M$ and $\tilde M $ can be arranged into orbits of $\varGamma_0 [2]$  and picking-up a single representative for each orbit allows one to disentangle the $\tau_1$ and $\tau_2$ integrals \KiritsisHF. In particular, one has to distinguish between the degenerate orbit
$$
M= \left( \matrix{ 0 &  \ldots & 0 \cr
2\ell_1 +1 &  \ldots & 2\ell_{2d}+1 \cr}\right)\,,
$$
and the non-degenerate one
$$
M= \left( \matrix{ 2n_1 &  \ldots & 2n_j & 0 & \ldots & 0 \cr
2\ell_1 +1 &  \ldots & 2\ell_j +1 & 2\ell_{j+1} +1 &  2\ell_{2d}+1 \cr}\right)\,,
\quad {\rm with}\qquad 
 2n_j > 2\ell_j +1 > 0 \,,
$$
and similarly for $\tilde M$. Let us consider, for instance, the contribution of the degenerate orbit to ${\scr V}_{\rm large}$. One finds
$$
\eqalign{
{\scr V}_{\rm large}^{\,\,\,\,({\rm deg})} =& - {\prod_i R_i \over 2(4\pi^2 )^{5-d} (\alpha ')^5} \int_{\sscr S} {d^2\tau\over \tau_2^{6}} \sum_{N=0}^\infty \sum_{N' = 0}^\infty\, \sum_{\{\ell\}} d_N d_{N'} e^{2i\pi(N-N') \tau_1 }
\cr 
&\qquad \qquad\times e^{-2\pi (N+N')\tau_2 -{\pi\over 4\alpha '\tau_2} \| (2\ell + 1)R\|^2}
\cr
=&- {\prod_i R_i \over 2(4\pi^2 )^{5-d} (\alpha ')^5} \int_0^\infty 
{d\tau_2\over \tau_2^{6}} \sum_{N=0}^\infty \sum_{\{\ell\}} d_N^2
e^{-4\pi N \tau_2 -{\pi\over 4\alpha '\tau_2} \| (2\ell + 1)R\|^2}
\cr}
$$
The above expression, however, cannot be computed analytically for arbitrary radius, since for values of the radius of the order of the string scale the $N$-series is not uniformly convergent and thus one cannot exchange the integration with the summation. However, one can easily extract the large-radius behaviour 
$$
{\scr V}_{\rm large}^{\,\,\,\,({\rm deg})} \sim - {1 \over R^{10-2d}} + O(e^{-R/\sqrt{\alpha'}})
$$
where, for simplicity we assumed the radii to be equal. The $R\to\infty$ behaviour of the non-degenerate orbit can be also computed to find the exponentially suppressed contributions of massive states. Similarly, had we started from ${\scr V}_{\rm small}$ the series would have been convergent only for small values of $R$, and 
$$
{\scr V}_{\rm small} \sim - R^{10-2d} + O(e^{-\sqrt{\alpha'}/R})\,, \qquad {\rm for}\quad R\to 0\,,
$$
whose behaviour could have been anticipated by T-duality arguments.

\vbox{
\vskip 10pt
\epsfxsize 12truecm
\centerline{\epsffile{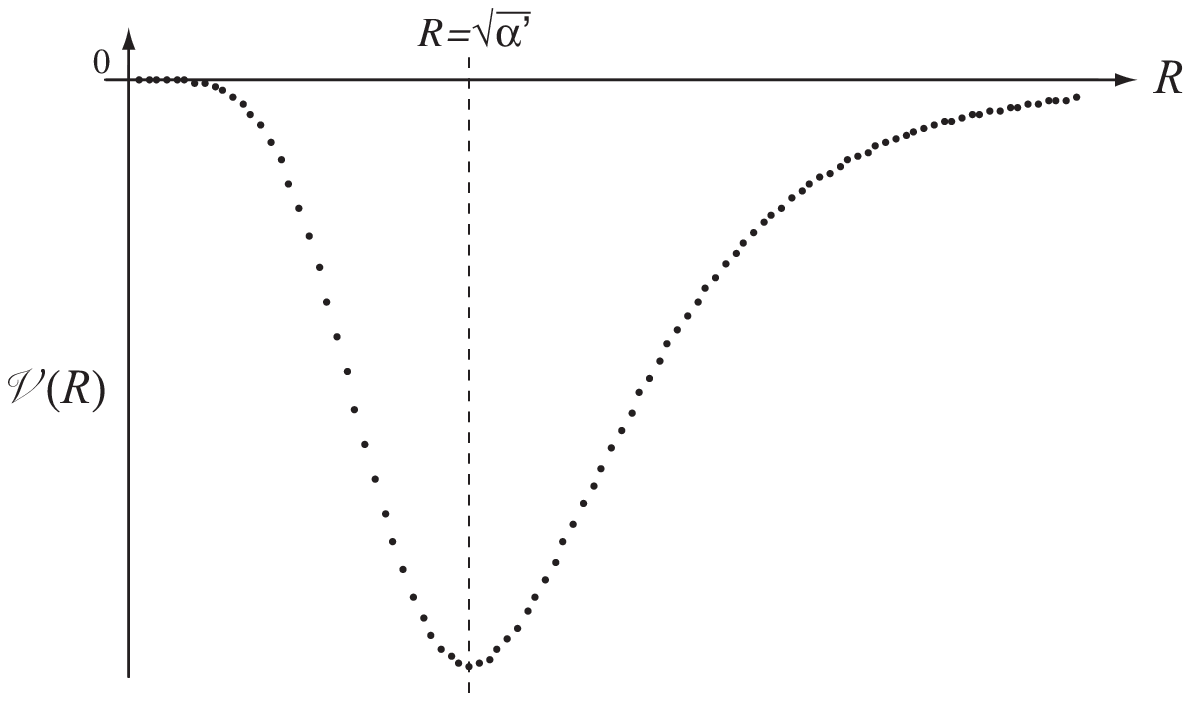}}
\vskip 10pt
\centerline{\ninepoint {\bf Figure.} Numerical evaluation of the one-loop potential as a function of the radius.}
\vskip 10pt
}

What about the behaviour of the potential for finite values of $R$? In the absence of tachyonic excitations no divergences are expected, while from the asymptotic behaviours
something special is expected to occur at the self-dual radius. Indeed, the gradient of the potential 
$$
{\partial {\scr V} \over \partial R_i } = {\pi \alpha' \over 2(4\pi^2 \alpha')^{5-d} } \int_{\tilde{\!\sscr F}} {d^2\tau\over \tau_2^{5-d}} \, \left| {\theta_2^4 \over \eta^{12}}\right|^2 \sum_{\{m,n\} }  (-1)^{(m+n)\cdot\epsilon}\, \varLambda_{m,n} 
\sum_{i=1}^{2d} \left( - {m_i^2\over R_i^3} + {n_i^2 R_i \over (\alpha ')^2}\right) 
\eqno(3)
$$
vanishes for $R_i =\sqrt{\alpha '}$. Hence, ${\scr V}\ $ has a global minimum at the self-dual point, where the radii of the internal torus are stabilised. These results can be directly confirmed by a numerical evaluation of the integral as reported in figure in the case of a squared lattice.

In the two asymptotic $R_i\to 0$ and $R_i\to \infty$ regions the vacuum energy vanishes as a result of a full restoration of supersymmetry on a flat Minkowski background. The global minimum, instead, induces a negative vacuum energy, and it is not clear whether supersymmetry is actually restored on this anti-de-Sitter background. 

In the expressions above we have considered, for simplicity, the case of a squared torus, where the off-diagonal components of the metric and the NS-NS $B$-field have been set to zero. While our results can be easily generalised to include all geometric moduli, it is not clear, however, whether in this simple toroidal setting twisted tachyonic modes are always absent. Actually, whenever supersymmetry-breaking is induced by the space-time fermion index it has been shown that different Scherk-Schwarz deformations live in the same moduli space \KiritsisEN. Hence, we expect that non-vanishing vev's for the compact moduli might lead to unwanted instabilities. Although it is not clear whether an energy barrier separates our minimum from the unstable regions, it is nevertheless possible to concoct models where the ``dangerous directions'' of the moduli space are simply removed by some orbifold or orientifold projection. In turn, this has the natural consequence of allowing for four-dimensional chirality and non-Abelian gauge interactions, the main ingredients of the Standard Model of Particle Physics. 
As an example, a (would-be supersymmetric) $\bb{Z}_4 \times \bb{Z}_4$ orbifold, with each $\bb{Z}_4$ acting on a $T^4\subset T^6$, puts the metric in a diagonal form while the world-sheet parity $\varOmega$ projects away the $B$-field. As a result, only three radial directions survive, precisely as in our previous discussions. Clearly, one has to be sure that the additional contributions to the vacuum energy do not spoil the structure of the one-loop potential. In fact, both for the orbifold and orientifold actions no additional contributions are generated. On the one hand, in the $\bb{Z}_4 \times \bb{Z}_4$ orbifold only the main orbit is present and the projected and/or twisted amplitudes are vanishing identically since the generators preserve, in principle, supersymmetry. On the other hand, the sector with reversed GSO projection is not left-right symmetric, henceforth it cannot contribute to the Klein-bottle, annulus and M\"obius strip amplitudes that, in turn, do not yield additional (tree-level and one-loop) contributions if all tadpoles are properly cancelled. 

Clearly, it would be interesting to generalise this construction to more general supersymmetry-breaking set-ups. Extrema of the potential should then occur at points of enhanced symmetry and eventually yield a more complex landscape of (meta-)stable vacua. 
It might also be rewarding to combine this mechanism with brane-supersymmetry-breaking \refs{\SugimotoTX,\Ads,\AngelantonjJH,\Alda} where NS-NS tadpoles are typically uncancelled, thus introducing a new positive source for the potential. Barring the stabilisation of the dilaton, the combination of these two effects might uplift the AdS minimum to a de Sitter metastable vacuum, as in \KachruAW. 
Of course, a different question is whether the dilaton (which cannot be stabilised by our method) stays small in the AdS minimum, and whether higher-loop corrections might destroy the extrema of ${\scr V}$, but a definite answer is out of reach from present technology. 

To conclude, it is tempting to speculate about a finite-temperature interpretation of this new Scherk-Schwarz deformation. To start with, let us quickly review what is known for the standard $(-1)^F\, \delta_1$ deformation \refs{\AtickSI,\AntoniadisKH}. In Field Theory, the thermal ensemble at temperature $T$ can be conveniently studied by considering the propagation of fields on $\bb{R}^{d-1} \times S^1$, where $S^1$ has circumference $2\pi\beta$. Indeed, at the one-loop level, also the free energy of a thermal gas of superstrings can likewise be computed \refs{\PolchinskiZF,\DitsasPM}
by considering the propagation over $\bb{R}^{d-1} \times S^1$ modded-out by $(-1)^F \,\delta_\beta$. Obviously, the superstring spectrum can be obtained from eq. (2) by identifying the temperature $T$ with the inverse radius $R$. In particular, the most (super)symmetric configuration at infinite $R$ corresponds to the most ordered phase at zero temperature, while in both cases a first-order phase transition occurs at some finite value of $R$ and/or $T$, as observed by Hagedorn. What could then be the thermodynamics interpretation of the new $(-1)^F \, \delta_3$ deformation we have now employed?
One of its main finger-prints is the large-radius--small-radius duality together with the absence of a first-order phase transition. This clearly leads to an analogy with the two-dimensional Ising model, where the Kramers-Wannier duality relates the ferromagnetic phase at low temperature to the paramagnetic phase at high temperature. Moreover, from the expression (3) one can deduce that the second (third) derivative is logarithmically divergent for $d=3$ ($d=2$), and thus a higher-order phase transition might take place. Again, this is analogous to the two-dimensional Ising model where a second-order phase transition takes place at the self-dual temperature, which can be smoothly crossed by continuous variations of the temperature. Clearly, it would be interesting to better investigate the thermodynamics interpretation of our model and relate it to the new phase transitions that seem to emerge in (non-)critical strings \refs{\OsorioHI{--}
\DienesXA}.

\vskip 0.5truein
\noindent
{\bf Acknowledgements.} C.A. and N.I. are pleased to thank Ignatios Antoniadis, Keith Dienes, Emilian Dudas, Elias Kiritsis, Francesco Knechtli, Costas Kounnas, Domenico Seminara, Stephan Stieberger, Tom Taylor and Miguel Vazquez-Mozo for useful discussions. 
M.C.  would like  to thank Giovanni Arcioni,  Shmuel Elitzur, Federico Elmetti, Alon Faraggi, 
Johannes Grosse,  Andrea Mauri, Bortolo Mognetti, Alessandro Nigro, Alberto Santambrogio, Leonard Susskind and  Cristina Timirgaziu  for illuminating discussions.
C.A. thanks the Physics Department of the University of Crete, the
 Gelileo Galilei Institute for Theoretical Physics and the Theory Unit at Cern for hospitality during the completion of this work. M.C. thanks the Department of Theoretical Physics of the University of Turin, the Racah Institute at the Hebrew University in Jerusalem, the Theory Unit at CERN, and The Theoretical Physics Department of the Liverpool University for hospitality.
N.I. thanks the Department of Theoretical Physics of the University of Turin, the Gelileo Galilei Institute for Theoretical Physics and the Theory Unit at Cern for hospitality. 
This work was supported in part by the European Community's Human Potential Programme under the contract HPRN-CT-2004-005104 to which both the University of Turin and the University of Milan 1 belong, in part by the European Community's Human Potential Programme under the contract HPRN-CT-2004-512194 to which the University of Crete belongs, and in part by INFN. The work of M.C. is partially supported also by PRIN prot.2005024045-002, and by  the  Royal Society grant n. RGMFHB.

{\ninepoint
\listrefs
}
\end